\documentclass[twocolumn,showpacs,preprintnumbers,amsmath,amssymb]{revtex4}

\usepackage{graphicx}
\usepackage{dcolumn}
\usepackage{bm}
\usepackage{epsfig}
\usepackage{amssymb}
\usepackage{amsthm}
\usepackage{verbatim}
\usepackage{delarray}
\usepackage{graphics}
\usepackage{epic}

\newcommand{\ket}[1]{{|#1\rangle}}
\newcommand{\bra}[1]{{\langle#1|}}
\newcommand{\braket}[2]{{\langle#1|#2\rangle}}

\newcommand{\tr}{\mbox{Tr}}
\newcommand{\rank}{\mbox{rank}}

\newcommand{\hpi}{\widehat{\Pi}}
\newcommand{\hx}{\widehat{X}}

\newcommand{\B}{{\mathcal{B}}}

\newcommand{\HH}{{\mathcal{H}}}

\newcommand{\SSS}{{\mathcal{S}}}

\newcommand{\ie}{{\em i.e., }}

\newtheorem{theorem}{Theorem}

\begin{document}

\title{Von Neumann Measurement is Optimal for Detecting Linearly Independent Mixed Quantum States}
\author{Yonina C. Eldar}
\homepage{http://www.ee.technion.ac.il/Sites/People/YoninaEldar/}
\email{yonina@ee.technion.ac.il} \affiliation{ Technion---Israel
Institute of Technology, Technion City, Haifa 32000, Israel }

\date{\today}

\begin{abstract}
We consider the problem of designing a measurement to minimize the
probability of a detection error when distinguishing  between a
collection of possibly non-orthogonal mixed quantum states. We
show that if the quantum state ensemble consists of linearly
independent density operators then the optimal measurement is an
orthogonal Von Neumann measurement consisting of mutually
orthogonal projection operators and not a more general positive
operator-valued measure.

\end{abstract}

\maketitle

\section{Introduction}

One of the important features of quantum mechanics is that non
orthogonal quantum states cannot be perfectly distinguished
\cite{P95}. Therefore, a fundamental problem in quantum mechanics
is to design measurements optimized to distinguish between a
collection of nonorthogonal quantum states.

We consider a quantum state ensemble consisting of $m$ density
operators $\{\rho_i, 1 \le i \le m\}$ on an $n$-dimensional
complex Hilbert space $\HH$, with prior probabilities $\{p_i>0, 1
\le i \le m\}$. A density operator $\rho$ is a positive
semidefinite (PSD)
 Hermitian operator with $\tr(\rho)=1$; we write $\rho \geq 0$ to
indicate $\rho$ is PSD. A mixed state ensemble is one in which at
least one of the density operators $\rho_i$ has rank larger than
one. A pure-state ensemble is one in which each density operator
$\rho_i$ is a rank-one projector $\ket{\phi_i}\bra{\phi_i}$, where
the vectors $\ket{\phi_i}$, though evidently normalized to unit
length, are
 not necessarily orthogonal.

In a quantum detection problem a transmitter conveys classical
information to a receiver using a quantum-mechanical channel. Each
message is represented by preparing the quantum channel in one of
the ensemble states $\rho_i$. At the receiver, the information is
detected by subjecting the channel to a quantum measurement in
order to determine the state prepared. If the quantum states are
mutually orthogonal, then the state can be determined correctly
with probability one by performing an optimal Von Neumann
measurement \cite{P95} consisting of $m$ mutually orthogonal
projection operators $\{\Pi_i\Pi_j=\delta_{ij}\Pi_i, 1 \le i,j \le
m\}$ that form a resolution of the identity on $\HH$ so that
$\sum_{i=1}^m\Pi_i=I$.

If the given states are not orthogonal, then no measurement will
distinguish perfectly between them.  Our  problem is therefore to
construct a measurement that minimizes the probability of a
detection error. It is well known that the most efficient way of
obtaining information about the state of a quantum system is not
always by performing orthogonal projections \cite{P90,PW91}, but
rather by performing more general positive operator-valued
measures (POVMs). A POVM consists of $m$ PSD Hermitian operators
$\{\Pi_i, 1 \le i \le m\}$ that form a resolution of the identity
on $\HH$ but are not constrained to be projection operators.

Necessary and sufficient conditions for an optimum measurement
maximizing the probability of correct detection have been
developed \cite{H73,YKL75,EMV02}. However, in general obtaining a
closed form expression for the optimal measurement directly from
these conditions is a difficult and unsolved problem. Closed-form
analytical expressions for the optimal measurement have been
derived for several special cases
\cite{H76,CBH89,OBH96,BKMO97,EF01,EMV02s}.

Kennedy \cite{K73} showed that for a pure state ensemble with
linearly independent vectors $\ket{\phi_i}$ the optimal
measurement is a Von Neumann measurement consisting of mutually
orthogonal rank-one projection operators. However, this
o,plication has not been proven for the more general case of {\em
mixed} state ensembles.

In Section~\ref{sec:von} we show that the optimal measurement for
distinguishing between a set of linearly independent mixed quantum
states is a Von Neumann measurement and not a general POVM.
Therefore, when seeking the optimal measurement, we may restrict
our attention to the class of Von Neumann measurements. We also
show that the rank of each projection operator is equal to the
rank of the corresponding density operator.

In Section~\ref{sec:lsm} we consider the least-squares measurement
(LSM) \cite{EF01}, also known as the square-root measurement
\cite{HW94,H96}, which is a detection measurement that has many
desirable properties and has been employed in many settings. We
show that for linearly independent mixed state ensembles the LSM
reduces to a Von Neumann measurement.

In the next section we present our detection problem and summarize
results from \cite{EMV02} pertaining to the conditions on the
optimal measurement operators.

\section{Optimal Detection of Quantum States}
\label{sec:qd}

Assume that a quantum channel is prepared in a quantum state drawn
from a collection of given states represented by density operators
$\{ \rho_i,1 \leq i \leq m \}$ in  an $n$-dimensional complex
Hilbert space $\HH$. We assume without loss of generality that the
eigenvectors of $\rho_i,1 \leq i \leq m$, collectively
span\footnote{Otherwise we can transform the problem to a problem
equivalent to the one considered here by reformulating the problem
on the subspace spanned by the eigenvectors of $\{\rho_i,1 \leq i
\leq m\}$. } $\HH$.

Since each density operator $\rho_i$ is Hermitian and PSD, it can
be expressed via the eigendecomposition as $\rho_i=\phi_i\phi_i^*$
where $\phi_i$ is an $n \times r_i$ matrix of orthogonal
eigenvectors $\{\ket{\phi_{ik}},1 \leq k \leq r_i\}$ and
$r_i=\rank(\rho_i)$. The density operators $\rho_i,1 \leq i \leq
m$ are {\em linearly independent} if the eigenvectors
$\{\ket{\phi_{ik}},1 \leq k \leq r_i,1 \leq i \leq m\}$ form a
linearly independent set of vectors. Since the eigenvectors of
$\rho_i,1 \leq i \leq m$ collectively span the $n$-dimensional
space $\HH$, it follows that for linearly independent state sets
\begin{equation}
\label{eq:sri} \sum_{i=1}^m r_i=n.
\end{equation}

At the receiver, the constructed measurement comprises $m$
measurement operators $\{\Pi_i,1 \leq i \leq m\}$ on $\HH$ that
satisfy
\begin{eqnarray}
\label{eq:identu}
\Pi_i & \geq & 0,\quad 1 \leq i \leq m;\nonumber \\
\sum_{i=1}^m \Pi_i & = & I_n,
\end{eqnarray}
where $I_n$ is the identity operator on $\HH$. We seek the
measurement operators $\{\Pi_i,1 \leq i \leq m\}$ satisfying
(\ref{eq:identu}) that maximize the probability of correct
detection which is given by
\begin{equation}
\label{eq:pe} P_d=\sum_{i=1}^mp_i\tr(\rho_i\Pi_i),
\end{equation}
where $p_i>0$ is the prior probability of $\rho_i$, with $\sum_i
p_i=1$.

It was shown in \cite{YKL75,EMV02} that a set of measurement
operators $\{\hpi_i,1 \leq i \leq m\}$ maximizes the probability
of correct detection for a state set $\{\rho_i,1 \leq i \leq m\}$
with prior probabilities $\{p_i,1 \leq i \leq m\}$ if and only if
there exists an Hermitian $\hx$ satisfying
\begin{equation}
\label{eq:condhx} \hx \geq p_i \rho_i,\quad 1 \leq i \leq m,
\end{equation}
such that
\begin{equation}
\label{eq:condz} (\hx-p_i\rho_i)\hpi_i=0,\quad 1 \leq i \leq m.
\end{equation}
The matrix $\hx$ can be determined as the solution to the problem
\begin{equation}
\label{eq:min} \min_{X \in \B} \tr(X)
\end{equation}
where $\B$ is the set of Hermitian operators on $\HH$, subject to
\begin{equation}
\label{eq:condx} X \geq p_i\rho_i,\quad 1 \leq i \leq m.
\end{equation}
As shown in \cite{EMV02}, the conditions (\ref{eq:condhx}) and
(\ref{eq:condz}) together imply that
\begin{equation}
\label{eq:tr} t_i \leq r_i,
\end{equation}
where $t_i=\rank(\hpi_i)$.

Kennedy \cite{K73} showed that for pure state ensembles
$\rho_i=\ket{\phi_i}\bra{\phi_i}$ with linearly independent
vectors $\ket{\phi_i}$ the optimal measurement is a rank-one
measurement $\hpi_i=\ket{\mu_i}\bra{\mu_i}$ with orthonormal
vectors $\ket{\mu_i}$, \ie a Von Neumann measurement. However,
this implication has not been proven for mixed states. In the
following section we use the conditions for optimality to prove
that the optimal measurement for linearly independent mixed states
is a Von Neumann measurement and not a more general POVM.

\section{Linearly Independent State Ensembles}
\label{sec:von}

Suppose now that the density operators $\rho_i$ are linearly
independent and let $\hpi_i$ be the optimal measurement operators
that maximize (\ref{eq:pe}) subject to (\ref{eq:identu}). Denoting
$\Pi=\sum_{i=1}^m \hpi_i$ we have that
\begin{equation}
\label{eq:ti1} \rank(\Pi) \leq \sum_{i=1}^m \rank(\hpi_i)=
\sum_{i=1}^m t_i.
\end{equation}
Since $\Pi=I_n$ we also have
\begin{equation}
\label{eq:ti2} \rank(\Pi) =n,
\end{equation}
from which we conclude that
\begin{equation}
\label{eq:tic} \sum_{i=1}^m t_i  \geq n.
\end{equation}
Combining (\ref{eq:tic}) with (\ref{eq:tr}) and (\ref{eq:sri}) we
conclude that
\begin{equation}
\label{eq:tic2} t_i=r_i.
\end{equation}
Therefore, via the eigendecomposition we can express each
measurement operator $\hpi_i$ as $\hpi_i=\mu_i\mu_i^*$ where
$\mu_i$ is an $n \times r_i$ matrix of orthogonal eigenvectors
$\{\ket{\mu_{ik}},1 \leq k \leq r_i\}$. Since $\sum_{i=1}^m r_i=n$
we have $n$ vectors $\ket{\mu_{ik}}$. In addition, from
(\ref{eq:identu}),
\begin{equation}
\label{eq:identu2} \sum_{ik}\ket{\mu_{ik}}\bra{\mu_{ik}}=I_n
\end{equation}
from which we conclude that the vectors $\{\ket{\mu_{ik}},1 \leq k
\leq r_i,1 \leq i \leq m\}$ are linearly independent.

We now show that the vectors $\{\ket{\mu_{ik}},1 \leq k \leq r_i,1
\leq i \leq m\}$ are mutually orthonormal. From (\ref{eq:identu})
we have that for any $1 \leq l \leq r_i,1 \leq j \leq m$,
\begin{equation}
\ket{\mu_{jl}}= \sum_{ik}
\braket{\mu_{ik}}{\mu_{jl}}\ket{\mu_{ik}}.
\end{equation}
Since the vectors $\ket{\mu_{ik}}$  are linearly independent, we
must have that $\braket{\mu_{ik}}{\mu_{jl}}=\delta_{ij,kl}$.

 We
conclude that
\begin{equation}
\hpi_i=\sum_{k=1}^{r_i} \ket{\mu_{ik}}\bra{\mu_{ik}}=P_{\SSS_i}
\end{equation}
where $P_{\SSS_i}$ is an orthogonal projection onto a subspace
$\SSS_i$ of  $\HH$ with dimension $r_i$ and
\begin{equation}
P_{\SSS_i}P_{\SSS_j}=\delta_{ij}P_{\SSS_i},
\end{equation}
so that $\HH=\SSS_1 \oplus \ldots \oplus \SSS_m$ is the direct sum
of the subspaces $\SSS_i$.

We summarize our results in the following theorem:
\begin{theorem}
\label{thm:von} Let $\{\rho_i,1 \leq i \leq m\}$ be a quantum
state ensemble consisting of linearly independent density
operators $\rho_i$ with prior probabilities $p_i>0$. Then the
optimal measurement is a Von Neumann measurement with measurement
operators $\{\hpi_i=P_{\SSS_i},1 \leq i \leq m\}$ where
$P_{\SSS_i}$ is an orthogonal projection onto an $r_i$-dimensional
subspace $\SSS_i$ of $\HH$ with $r_i=\rank(\rho_i)$ and
$P_{\SSS_i}P_{\SSS_j}=\delta_{ij}P_{\SSS_i}$.
\end{theorem}

\section{Least-Squares Measurement}
\label{sec:lsm}

A suboptimal measurement that has been employed as a detection
measurement in many applications is the least-squares measurement
(LSM) \cite{EF01,H98}, also known as the square-root measurement
\cite{BKMO97,HW94,H96,SKIH98,SUIH98,KOSH99}. The LSM has many
desirable properties. Its construction is relatively simple; it
can be determined directly from the given collection of states; it
minimizes the probability of a detection error for pure and mixed
state ensembles that exhibit certain symmetries \cite{EF01,EMV02};
it is ``pretty good''  when the states to be distinguished are
equally likely and almost orthogonal \cite{HW94};  and it is
asymptotically optimal \cite{H96,H98}.

The LSM corresponding to a set of density operators
$\{\rho_i=\phi_i\phi_i^*,1 \leq i \leq m\}$ with eigenvectors that
collectively span $\HH$ and prior probabilities $\{p_i,1 \leq i
\leq m\}$ consists of the measurement operators
$\{\Sigma_i=\mu_i\mu_i^*,1 \leq i \leq m\}$ where \cite{H98,EF01}
\begin{equation}
\label{eq:glsm} \mu_i=(\Psi\Psi^*)^{-1/2}\psi_i.
\end{equation}
Here $\Psi$ is the matrix of (block) columns
$\psi_i=\sqrt{p_i}\phi_i$ and $(\cdot)^{1/2}$ is the unique
Hermitian square root of the corresponding matrix. Note that since
the eigenvectors of the $\{\rho_i\}$ collectively span $\HH$, the
columns of the $\{\psi_i\}$ also together span $\HH$, so
$\Psi\Psi^*$ is invertible.

We now show that the LSM satisfies the conditions of
Theorem~\ref{thm:von} so that if the columns of $\{\phi_i\}$ are
linearly independent, then $\Sigma_i\Sigma_j=\Sigma_i\delta_{ij}$
and the LSM is a Von Neumann measurement.

From (\ref{eq:glsm}) we have that
\begin{equation}
\label{eq:sisj}
\Sigma_i\Sigma_j=(\Psi\Psi^*)^{-1/2}\psi_i\psi_i^*(\Psi\Psi^*)^{-1}
\psi_j\psi_j^*(\Psi\Psi^*)^{-1/2}.
\end{equation}
To simplify (\ref{eq:sisj}) we express $\psi_i$ as
\begin{equation}
\label{eq:psii} \psi_i=\Psi E_i.
\end{equation}
Here $E_i$ is an $n \times r_i$ matrix where the $q$th column of
$E_i$ has one nonzero element equal to $1$ in the $p$th position
with $p=\sum_{k=1}^{i-1}r_k+q$. We then have that
\begin{equation}
 \psi_i^*(\Psi\Psi^*)^{-1} \psi_j=
E_i^*\Psi^*(\Psi\Psi^*)^{-1} \Psi E_j.
\end{equation}
If the density operators $\{\rho_i\}$ are linearly independent,
then $\sum_i r_i=n$ and the operators $\{\psi_i\}$ are also
linearly independent. Since each matrix $\psi_i$ has dimension $n
\times r_i$ we conclude that $\Psi$ is a $n \times n$ matrix with
linearly independent columns and therefore invertible. Thus,
$\Psi^*(\Psi\Psi^*)^{-1} \Psi=I_n$ and
\begin{equation}
\label{eq:sisjt} \psi_i^*(\Psi\Psi^*)^{-1} \psi_j=
E_i^*E_j=\delta_{ij}I.
\end{equation}
Substituting (\ref{eq:sisjt}) into (\ref{eq:sisj}),
\begin{equation}
\Sigma_i\Sigma_j=\delta_{ij}(\Psi\Psi^*)^{-1/2}\psi_i\psi_i^*(\Psi\Psi^*)^{-1/2}=
\delta_{ij}\Sigma_i
\end{equation}
and the LSM is a Von Neumann measurement consisting of mutual
orthogonal projection operators.

We summarize our results regarding the LSM in the following
theorem.
\begin{theorem}
\label{thm:lsm} Let $\{\rho_i,1 \leq i \leq m\}$ be a quantum
state ensemble consisting of linearly independent density
operators $\rho_i$ with prior probabilities $p_i>0$. Then the
least-squares measurement is a Von Neumann measurement with
measurement operators $\{\Sigma_i=P_{\SSS_i},1 \leq i \leq m\}$
where $P_{\SSS_i}$ is an orthogonal projection onto an
$r_i$-dimensional subspace $\SSS_i$ of $\HH$ with
$r_i=\rank(\rho_i)$ and
$P_{\SSS_i}P_{\SSS_j}=\delta_{ij}P_{\SSS_i}$.
\end{theorem}

\begin{acknowledgments}

The author is a Horev Fellow supported by the Taub Foundation.

\end{acknowledgments}


\end{document}